\documentclass[iop]{emulateapj}
\usepackage{amssymb}
\usepackage{amsmath}
\shorttitle{Milliarcsecond radio sources } \shortauthors{Cao et al.}

\begin{document}
\title{Exploring the properties of milliarcsecond radio sources }
\author{ Shuo Cao \altaffilmark{1}, Marek Biesiada\altaffilmark{1,2}, Xiaogang Zheng\altaffilmark{1}, and Zong-Hong Zhu\altaffilmark{1*}
}

\altaffiltext{1}{Department of Astronomy, Beijing Normal University,
100875, Beijing, China; \emph{zhuzh@bnu.edu.cn}}
\altaffiltext{2}{Department of Astrophysics and Cosmology, Institute
of Physics, University of Silesia, Uniwersytecka 4, 40-007 Katowice,
Poland}

\begin{abstract}
Cosmological applications of the ``redshift - angular size" test
require knowledge of the linear size of the "standard rod" used. In
this paper, we study the properties of a large sample of 140
milliarcsecond compact radio sources with flux densities measured at
6 cm and 20 cm, compiled by Gurvits et al.(1999). Using the
best-fitted cosmological parameters given by Planck/WMAP9
observations, we investigate the characteristic length $l_m$ as well
as its dependence on the source luminosity $L$ and redshift  $l_m=l
L^\beta (1+z)^n$. For the full sample, measurements of the
angular size $\theta$ provide a tight constraint on the linear size
parameters. We find that cosmological evolution of the linear size
is small ($|n|\simeq 10^{-2}$) and consistent with previous
analysis. However, a substantial evolution of linear sizes with
luminosity is still required ($\beta\simeq 0.17$). Furthermore,
similar analysis done on sub-samples defined by different source
optical counterparts and different redshift ranges, seems to support
the scheme of treating radio galaxies and quasars with distinct
strategies. Finally, a cosmological-model-independent method is
discussed to probe the properties of angular size of milliarcsecond
radio quasars. Using the corrected redshift - angular size relation
for quasar sample, we obtained a value of the matter density
parameter, $\Omega_m=0.292^{+0.065}_{-0.090}$, in the spatially flat
$\Lambda$CDM cosmology.

\end{abstract}


\section{Introduction}\label{sec:introduction}

The redshift - angular size data have provided a useful
method to probe cosmological parameters
\citep{Guerra00,Vishwakarma01,Lima02,Chen03}, since this relation is directly related to the angular diameter
distance. Powerful radio sources constitute a population which
can be observed up to very high redshifts, reaching beyond
feasible limits of supernova studies. Over a few past decades considerable advances have
been made to investigate the redshift -
angular size relation in radio sources for the purpose of cosmological studies,
 including the works of
\citep{Singal93,Daly94,Kayser95,Buchalter98,Gurvits99,Guerra00,Zhu02,Podariu03,Jackson04,Barai06,Barai07}.
Up to now, redshift - angular size relation has been measured for
different types of radio sources, such as extended FRIIb galaxies
\citep{Daly03}, radio loud quasars \citep{Buchalter98} and radio
galaxies \citep{Guerra00}. In a similar spirit, by using radio
observations of the Sunyaev-Zeldovich effect (SZE) together with
X-ray emission of galaxy clusters, \citet{Filippis05,Boname06}
extensively explored the angular diameter distances at different
redshifts.

In fact, in order to break inherent
degeneracies between cosmological parameters every alternative method of restricting
these parameters is desired in modern cosmology.
Consequently, there have been numerous attempts to use compact radio sources for this purpose
\citep{Vishwakarma01,Lima02,Zhu02,Chen03}. In these studies the
analysis was carried out on 12 binned data-points with 12-13 compact
sources per bin. One of the major uncertainties was the typical value of the
linear size $l$. In order to obtain cosmological constraints, some authors chose
to fix $l$ at certain specific values
\citep{Vishwakarma01,Lima02,Zhu02}, while \citet{Chen03} chose to
include some range of values for $l$ and then marginalized over it.

It is obvious that cosmological application of the ``redshift -
angular size'' data requires good knowledge of the linear size of
the ``standard rod'' used. The possibility that source's linear size
depends on the source luminosity and redshift should be kept in
mind. In particular it still remains controversial whether compact
radio sources are indeed ``true'' standard rods
\citep{Gurvits99,Vishwakarma01}. Applying the popular
parametrization $l_m=lL^\beta(1+z)^n$, \citet{Gurvits99} claimed
that by excluding sources with extreme spectral indices and low
luminosities their compiled data has been minimized for the
confounding by these two effects \citep{Gurvits99,Vishwakarma01}.
However, their results were obtained under assumption of a
homogeneous, isotropic universe without cosmological constant
($\Omega_\Lambda=0$). On the other hand, from a large body of recent
astronomical observations, such as Union2 SNe Ia dataset
\citep{Amanullah10}, the CMB observation from the Wilkinson
Microwave Anisotropy Probe (WMAP9) \citep{Hinshaw12}, and the BAO
distance ratios from the spectroscopic Sloan Digital Sky Survey
(SDSS) data release7 (DR7) galaxy sample \citep{Padmanabhan12}, no
convincing evidence for deviations from the concordance $\Lambda$CDM
model has been established. More recently, Planck, the
third-generation space mission following COBE and WMAP, has recently
released its first cosmological results based on measurements of the
CMB temperature and lensing-potential power spectra \citep{Planck1}.
All of them strongly indicate the existence of an exotic component
called dark energy, which represents more than 70\% of the total
energy of the universe and serving as a driving force of the cosmic
acceleration. Latest investigations specifically to study the
properties of dark energy were carried out by
\citet{Yu11,Cao11c,Cao12a,Cao12b,Cao12c,Cao12d,Pan12,Cao13,Liao13,Cao14}.
Having this in mind, properties of compact radio sources should be
readdressed with present angular size data and taking into account a
reliable cosmology based on current precise observations.

In this paper, we will reconsider issues associated with angular
sizes of radio sources under the assumption of $\Lambda$CDM
cosmological model. Specifically, we will study the characteristic
length, the ``angular size - luminosity" and ``angular size -
redshift" relations for the compact structure in quasars and radio
galaxies assuming $\Lambda$CDM cosmological model as the background,
which is well supported by observations. In Section 2 and 3 we
briefly describe our sample, its construction and methodology of
subsequent analysis. Results with the full sample and several
sub-samples are presented in Section 4.
Cosmological-model-independent constraints on the compact source
parameters and their cosmological application are discussed in
Section 5. Finally, we summarize the conclusions in Section 6.

\begin{table*}
\caption{\label{tab:result} Summary of constraints on the metric linear size parameters obtained with
the full sample and six sub-samples (see text for definitions). Source of cosmological priors and the wavelength of measured flux are given in brackets.}
\begin{center}
\begin{tabular}{l|l|l|lllllllllllllll}\hline\hline
Sample (Cosmology+Flux density)        & $l$ (pc)  & $\beta$  &   $n$ \\
\hline
Full sample (Planck+$S_6$)  & $l= 25.42\pm 3.62$    & $\beta= 0.169\pm 0.025$  & $n=-0.021\pm 0.139$  \\
Full sample (WMAP9+$S_6$)   & $l= 24.91\pm 3.59$   & $\beta= 0.170\pm 0.026$  & $n=-0.009\pm 0.141$  \\
Full sample (Planck+$S_{20}$)  & $l= 27.07\pm 3.95$    & $\beta= 0.183\pm 0.027$  & $n=-0.090\pm 0.145$  \\

\hline

Radio galaxy (Planck+$S_6$)  & $l= 49.55^{+26.35}_{-20.05}$    & $\beta= 0.242\pm 0.064$  & $n= 0.142\pm 0.670$  \\

Quasar (Planck+$S_6$)  & $l= 25.96\pm 4.14$    & $\beta= 0.203\pm 0.034$  & $n= -0.023\pm 0.153$  \\

BL Lac (Planck+$S_6$)  & $l= 20.73^{+23.77}_{-11.78}$    & $\beta= 0.139\pm 0.107$  & $n= -0.315\pm 0.895$  \\
\hline

Sub-sample ($z\leq 0.5$) (Planck+$S_6$)  & $l= 44.60\pm21.80$    & $\beta= 0.284\pm 0.076$  & $n= 1.228\pm 1.072$  \\

Sub-sample ($0.5<z\leq 1.0$) (Planck+$S_6$)  & $l= 21.65^{+21.20}_{-13.20}$    & $\beta=0.009\pm 0.091$  & $n= -0.335\pm 1.115$  \\

Sub-sample ($1.0<z\leq 2.0$) (Planck+$S_6$)  & $l= 63.85^{+34.15}_{-31.85}$    & $\beta= 0.215\pm 0.083$  & $n= -0.880\pm 0.600$  \\

Sub-sample ($z>2.0$) (Planck+$S_6$)  & $l= 10.49^{+14.51}_{-8.34}$    & $\beta= 0.128\pm 0.187$  & $n= 0.518\pm 0.638$  \\

\hline
\end{tabular}
\end{center}
\end{table*}

\section{Observational data}\label{sec:data}

Our goal is to better constrain the parameters modeling compact
radio sources, i.e. their linear size scale $l$, along with
luminosity and redshift dependence of their metric linear length
$l_m$. By ``better'' we mean obtained using the best currently
available cosmological model. To this end, we have considered the
angular size data for milliarcsecond radio sources compiled by
\citet{Gurvits99}. This data set was a larger sample of sources than
used by \citet{Kellermann93} or by \citet{Wilkinson98} and with more
complete structural data than used by \citet{Gurvits93,Gurvits94}.

All 330 sources included in this comprehensive compilation were imaged
with very-long-baseline interferometry (VLBI) at 5 GHz with resolution of
ca. 1.5 mas. They included:  1) 79 compact radio
sources associated with active galaxies and quasars considered in
\citet{Kellermann93}; 2) sources described by the Caltech - Jodrell Bank group
\citep{Xu95,Henstock95,Taylor94,Taylor96}; 3) sources discussed in
other works as well as observations of high-redshift ($z>3$) quasars
\citep{Frey97,Paragi98}. \citet{Gurvits99} then reduced this original
compilation of 330 sources down to 145 sources, with spectral index ($-0.38\leq
\alpha\leq 0.18$) and total luminosity ($Lh^2\geq 10^{26}
WHz^{-1}$). They claimed that the former criterion helps excluding
compact sources with inverted spectrum and relatively large
steep spectrum, while the latter tends to minimize the possible
dependence of linear size on luminosity \citep{Gurvits99}.

Full information about all the 145 sources that remain after the
aforementioned selection, can be found in Table 1 of
\citet{Gurvits99}, including source coordinates, redshifts, optical
counterpart, angular size, spectral index, and total flux densities
at 6 cm or 20 cm. Following \citet{Kellermann93}, the characteristic
angular size of each source was defined as the distance between the
strongest component (referred to as the core) and the most distant
component which had the peak brightness greater than or equal to 2\%
of the peak brightness of the core. We emphasize here, that in order
to implement multi-frequency ``$\theta$-z" tests in our analysis
(see the next section for details), we further restricted the final
sample to 140 data points with measured flux densities at both 6 cm
and 20 cm. The final sample, which covers the redshift range
$0.031<z<3.89$ and does not show jet-like structure for any system,
contains a wide class of extragalactic objects including 112 sources
identified as quasars, 18 radio galaxies, and 10 BL Lac objects
(blazars).

Radio galaxies we use in this work are located within the redshift
range of $0<z<0.9$. This means that, using the ``redshift - angular
size" test, we are able to constrain properties of active galaxies
and their evolution up to $z\sim 0.9$. Main motivation of studying
the milliarcsecond radio structures in quasars stems from their
potential usefulness in cosmology \citep{Kellermann93,Gurvits99}.
Moreover, the large number of quasars is also beneficial for
studying structural properties of milliarcsecond radio structures at
high redshifts. Finally, inclusion of 10 sources with BL Lac objects
as counterparts allows us to show that their structural properties
are similar to the known quasars at $z>3$.

However, the ``angular size - redshift'' test requires statistically complete
and well-characterized (homogeneous) sample. Because our list includes sources
corresponding to different optical counterparts at different
redshift as described above, so besides the full combined sample we will also consider
separately seven sub-samples. Three of them are defined by optical counterparts: radio galaxies, quasars and blazars.
Next subsamples are defined by restriction to four redshift ranges:  $z\leq 0.5$, $0.5<z\leq 1.0$, $1.0<z\leq 2.0$ and $z > 2.0$.

\section{Method}\label{sec:method}

According to \citet{Sandage88}, the angular
size-redshift relation for a rod of intrinsic length $l_m$ can be
written as
\begin{equation}
\theta(z)= \frac{l_m}{D_A(z)} \label{theta}
\end{equation}
where $l_m$ is the metric linear size, $D_A$ is the angular diameter
distance at redshift $z$. Following the phenomenological model
first proposed in \citet{Gurvits94} and later discussed in
\citet{Gurvits99}, the projected linear size of a source is related
to its luminosity $L$ and redshift $z$ as
\begin{equation} \label{lm}
l_m=l(\frac{L}{L_0})^\beta(1+z)^n\,
\end{equation}
where $l$ is the linear size scaling factor representing the
apparent distribution of radio brightness from the peak down to the
level of its 2\% in the available sample of VLBI images. It is a
parameter defined by the practical limitation of dynamic range of
VLBI data, i.e., a higher sensitivity of VLBI observations would
enable estimates of angular sizes down to lower values of
brightness, resulting in turn in a different value of the
characteristic linear size \citep{Gurvits94}. $L_0$ is the
normalizing luminosity taken to be equal to $10^{28}$WHz$^{-1}$ in
our analysis. Parameters $\beta$ and $n$ represent the dependence of
the linear size on source luminosity and redshift, respectively. The
first parameter -- $\beta$ is related to physics of a compact radio
emitting regions, while the parameter $n$ mimics three physical
effects: (1) cosmological evolution of the linear size with
redshift; (2) dependence of the linear size on the emitted
frequency; and (3) an impact of sources broadening due to scattering
in the propagation medium \citep{Gurvits99}. The latter effect is
not important for our sample with the lowest emitted frequency of 5
GHz (corresponding to $z=0$). The distinction between the former two
effects require multi-frequency $\theta - z$ tests.

The luminosity $L$ of radio sources is estimated from their measured
flux density $S_{obs}$. So the radio luminosity,
assuming isotropic emission, reads:
\begin{equation}
L= \frac{S_{\mathrm{obs}} 4 \pi {D_{L}}^{2}}{(1+z)^{1+\alpha}}
\end{equation}
where $S_{obs}$ is the observed flux density, $D_L$ is the
luminosity distance, and $\alpha$ is the spectral index
($S_{\mathrm{obs}} \propto{\nu}^{\alpha}$). In general, for sources at cosmological
distances, k-correction must be applied to the spectral index
$\alpha$ of the source. The angular diameter
distance $D_A$ and the luminosity distance $D_L$ at redshift $z$ are
related to each other through the so-called distance duality
relation
\begin{equation}
D_{L}/D_A(1+z)^{-2}= 1
\end{equation}
which is a fundamental relation in observational cosmology
and initiated a lot of studies, e.g. \citep{Cao11a,Cao11b}.

The above equations imply that, if we could have a reliable knowledge of
cosmological model parameters which therefore allow to calculate
$D_A$ or $D_L $ at different redshifts, then we would get stringent constraints
on the range of parameters, $l$, $\beta$,
and $n$ describing compact radio sources.
Theoretical expression for the angular
diameter distance $D_A(z)$ (expressed in Mpc and assuming flat FRW metric) reads
\begin{eqnarray}
\label{inted} D_A(z;\Omega_m)=\frac{3000h^{-1}}{(1+z)}\int_{0}^{z}
\frac{dz'}{E(z';\Omega_m)}
\end{eqnarray}
where $h$ is the dimensionless Hubble constant, $E(z; \Omega_m)$ is
the dimensionless expansion rate, which -- in the case of flat $\Lambda$CDM model -- depends on redshift $z$ and
matter density parameter $\Omega_m$ in the following way:
\begin{equation}
E^2(z; \Omega_m)=\Omega_m(1+z)^3+(1-\Omega_m).
\end{equation}
For the purpose of our analysis, theoretical $D_A(z)$ has been
calculated by using the best-fit matter density parameter given by
Planck Collaboration: $\Omega_m=0.315\pm0.017$ and $h=0.673\pm0.012$
\citep{Planck1}. Even though Planck results are the latest ones, we
also include the data from the Wilkinson Microwave Anisotropy Probe
9 year data release (WMAP9), i.e. $\Omega_m=0.279\pm0.025$ and
$h=0.700\pm0.022$ \citep{Hinshaw12}. The $h=0.700$ value is also
used the cosmological application of the
cosmological-model-independent method is discussed in Section 5. In
order to determine the parameters of compact radio structures, we
preformed Monte Carlo simulations of the posterior likelihood ${\cal
L} \sim \exp{(- \chi^2 / 2)}$ using routines available in CosmoMC
package. As a prior, we assumed a conservative 20\% Gaussian
uncertainty of the observed angular size.

\section{Results and discussions}\label{sec:results}

In this paper, we focused our attention on the constraints on the
parameters ($l$, $\beta$, and $n$) characterizing compact radio sources
obtained from different samples,
i.e. the full $N=140$ sample as well as several sub-samples
determined from different selection criteria. The results are
summarized in Table 1.

\subsection{Estimates on the full sample}

As we already remarked, measured flux density at different bands
could bear the information about physical conditions in active
galactic nuclei -- a feature common to all types of sources we used.
Performing fits on the data comprising flux at 6 cm, we obtained the
following best-fit values and corresponding 1$\sigma$ (more
precisely 68\% confidence level) uncertainties
\begin{eqnarray}
&& l= 25.42\pm 3.62 \ \mathrm{pc}, \nonumber\\
&& \beta= 0.169\pm 0.025, \nonumber\\
&& n=-0.021\pm 0.139. \nonumber
\end{eqnarray}

\begin{figure}
\begin{center}
\includegraphics[width=1.0\hsize]{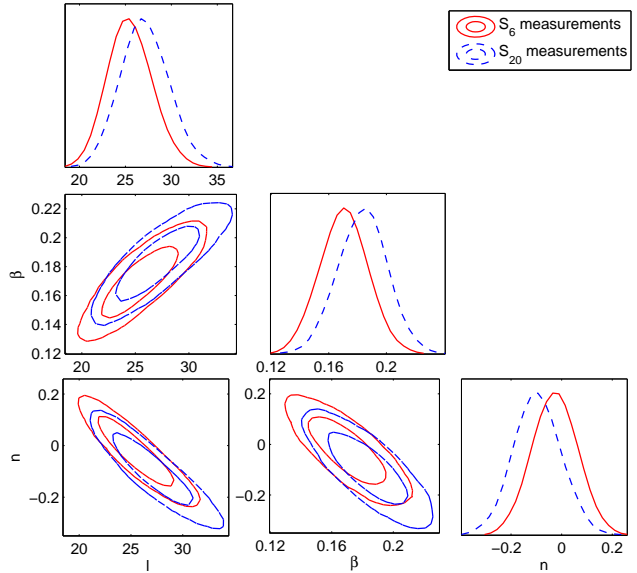}
\end{center}
\caption{Constraints on compact source parameters obtained from the
full sample, based on different flux density measurements at 6 cm
and 20 cm, respectively. \label{fig1}}
\end{figure}

Then, using the flux densities at 20 cm we obtain the following best
fit
\begin{eqnarray}
&& l= 27.07\pm 3.95 \ \mathrm{pc}, \nonumber\\
&& \beta= 0.183\pm 0.027, \nonumber\\
&& n=-0.090\pm 0.145. \nonumber
\end{eqnarray}
Marginalized 1$\sigma$ and 2$\sigma$ contours of each parameter
obtained at different bands are shown in Fig.~\ref{fig1}. It is
clear that the parameter degeneracies are consistent with each
other, as can be seen in the contours obtained from flux densities
measured at 6 cm and 20 cm, respectively. Best fitted values
obtained for different wavelength are different, but they agree
within 1$\sigma$.

It is obvious that, for well resolved compact sources, measurements
of $\theta$ provide tighter estimates of the linear size parameters
($l$, $\beta$, $n$). More importantly, our full sample analysis has
also yielded improved constraints on the meaningful physical
parameters: $\beta$ and $n$. We found that best-fitted value of the
parameter $n$ is a small number: $|n|\simeq 10^{-2}$, slightly
negative, but 68\% CI contains zero in any case i.e. our results are
consistent with no evolution of $l_m$ with $z$. This suggests that,
contrary to the case of extended radio sources, central engine
powering compact radio sources is likely to be controlled by a
limited number of physical parameters (mass of central black hole,
accretion rate) and may therefore be less subject to evolutionary
effects. On the other hand, we found that, for 140 sources
satisfying luminosity selection criterion $Lh^2\geq 10^{26}
WHz^{-1}$, substantial evolution of linear size with luminosity is
still required. Compared with previous results obtained on the same
data \citep{Gurvits99}, our results show that, improved, more
rigorous quantitative analysis supports the existence of ``linear
size - luminosity" relation. The conclusion that $\beta+n\geq-0.15$
given by \citet{Gurvits99} does not contradict to the findings of
the present work. The best-fit parameters of the phenomenological
dependence Eq.(\ref{lm}) under the modern cosmological model are
different from those obtained with the classical Einstein - de
Sitter model used by \citet{Gurvits99}. The values of the two
best-fit parameters of the phenomenological formula obtained here,
namely, $\beta$ and $n$, if confirmed by future "angular
size-redshift" studies, would offer additional constraints for
cosmological tests based on angular sizes of extragalactic sources.

\begin{figure}
\begin{center}
\includegraphics[width=1.0\hsize]{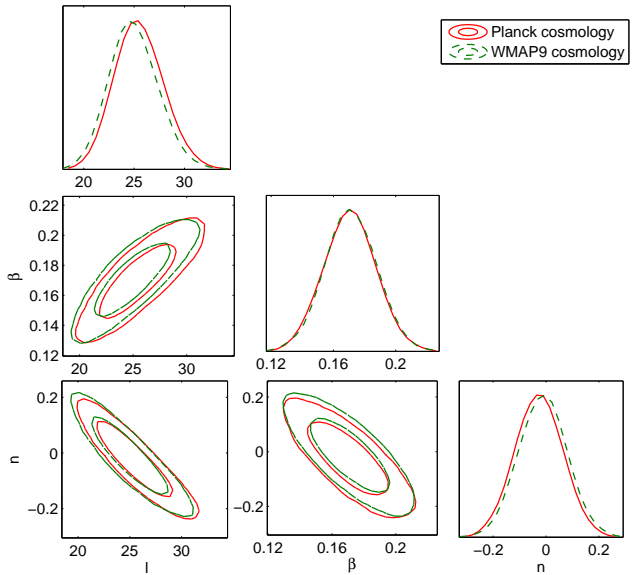}
\end{center}
\caption{Constraints on compact source parameters obtained from the
full sample, assuming cosmological parameters inferred from Planck
and WMAP9 data. \label{fig2}}
\end{figure}

As we stressed above, the assumption of currently best available cosmological  model
--- $\Lambda$CDM was the source of improvement concerning estimates of $l$, $\beta$, and $n$.
Hence their values depend on the
cosmological parameters used. Therefore, besides assuming flat $\Lambda$CDM
model with parameters coming from Planck observations, we
also considered WMAP9 results for comparison.
In this case, the best fit is
\begin{eqnarray}
&& l= 24.91\pm 3.59 \ \mathrm{pc}, \nonumber\\
&& \beta= 0.170\pm 0.026£¬\nonumber\\
&& n=-0.009\pm 0.141. \nonumber
\end{eqnarray}
Marginalized probability distributions for each parameter and
marginalized 2D 68\% confidence contours are presented in
Fig.~\ref{fig2}. Comparing constraints based on Planck and WMAP9
observations, we see that confidence regions of $l$, $\beta$, and
$n$ are almost the same, hence our results and discussions presented
above are robust. We remark here that, considering the WMAP9
and Planck data are consistent with the accuracy sufficient to the
comparison with the ``$\theta$-z" test, it is not surprising the
regression results of the ``$\theta$-z" test in combination with
WMAP and Planck are compatible in the framework of $\Lambda$CDM
cosmology.

\subsection{Estimates on sub-samples}

\begin{figure}
\begin{center}
\includegraphics[width=1.0\hsize]{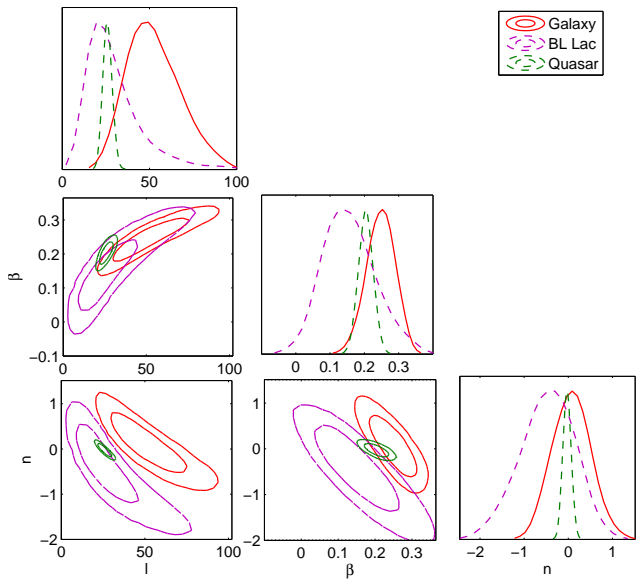}
\end{center}
\caption{Constraints on compact source parameters obtained from
three sub-samples with different optical counterparts. \label{fig3}}
\end{figure}

\begin{figure}
\begin{center}
\includegraphics[width=1.0\hsize]{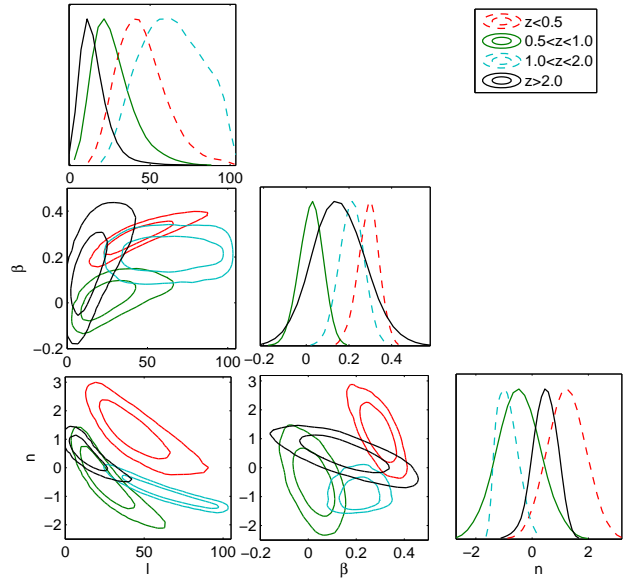}
\end{center}
\caption{Constraints on compact source parameters obtained from
sub-samples defined by four different redshift bins. \label{fig4}}
\end{figure}

In Table 1 and Fig.~\ref{fig3}-\ref{fig4}, we show the results of
fitting three parameters, $l$, $\beta$ and $n$ on seven sub-samples
described in Section~\ref{sec:data}.

We note that the ranges of $l$ and $\beta$ parameters for quasars
($l= 25.96\pm 4.14$ pc, $\beta= 0.203\pm 0.034$) are marginally
close to estimates obtained for compact structures in BL Lac objects
($l= 20.73^{+23.77}_{-11.78}$ pc, $\beta= 0.139\pm 0.107$). Rather
weak dependence of the characteristic size on redshift, i.e. the
range of the parameter $n$ for quasars ( $n= -0.023\pm 0.153$) is in
agreement with the estimate obtained for BL Lac sources ($n=
-0.315\pm 0.895$) within 1$\sigma$. On the other hand, luminosity
dependence ($\beta= 0.242\pm 0.064$) and weak redshift dependence
($n= 0.142\pm 0.670$) are both present in radio galaxies. The
best-fit values of $\beta$ and $n$ for this sub-population are
significantly different from the corresponding quantities of quasars
or BL Lac sources. Consequently, our results imply the existence of
physical differences between galaxies and quasars at the
milliarcsecond scale. To some extent, this conclusion supports the
scheme of treating radio galaxies and quasars with distinct
strategies. We must keep in mind that similarity or difference in
$(\beta, n)$ parameters for radio sources with different types of
optical counterparts, might reveal similar or different physical
processes governing the radio emission of compact structures.

This tendency could also be found in fits performed on four
sub-samples with different redshift bins. From Fig.~\ref{fig4} we
find that: (1) Constraints on all the parameters coming from the
low-redshit sub-sample ($z\leq 0.5$) are substantially different
from those obtained with other sub-samples. This can be explained by
the fact that that low redshift sub-sample is dominated by radio
galaxies. (2) For the sub-sample with redshift range $0.5<z\leq
1.0$, the ``no-evolution'' model ($\beta = n = 0$) is still included
within $1\sigma$ confidence regions in $(\beta,n)$ parameter plane,
whereas a substantial evolution of linear sizes with luminosity is
still required for the other three sub-samples.

As we remarked above, sub-samples defined by redshift ranges are confounded with types of optical counterparts.
Therefore a stratified analysis taking into account redshift ranges and the source type would be desirable. However,
our sample is too small to achieve this.

\begin{figure}
\begin{center}
\includegraphics[width=1.0\hsize]{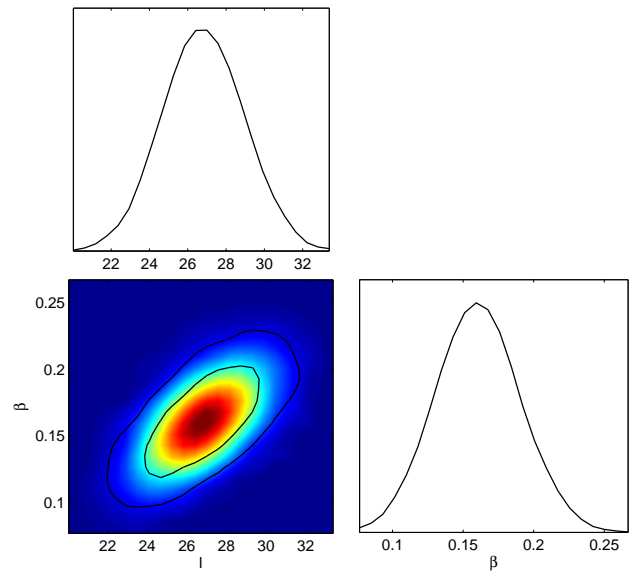}
\end{center}
\caption{Constraints from the $N=26$ quasar sub-sample, with
observational luminosity distance derived from Union2 SN Ia data.
\label{fig5}}
\end{figure}

\begin{figure}
\begin{center}
\includegraphics[width=1.0\hsize]{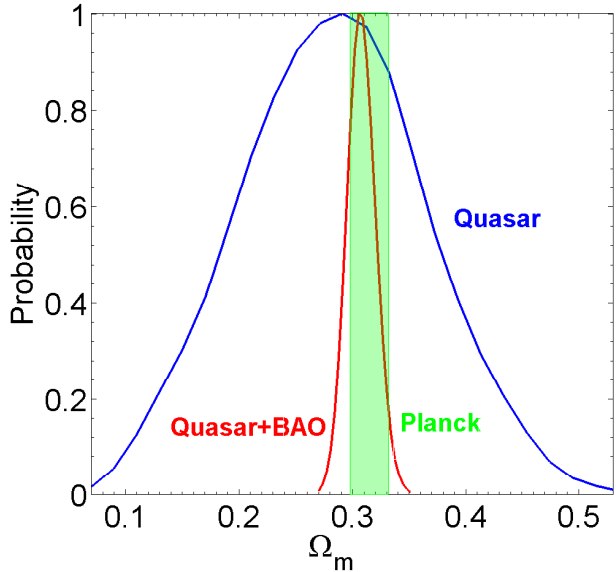}
\end{center}
\caption{Cosmological fits on the $\Lambda$CDM model form the
corrected angular size - redshift relation for quasars (Blue line)
and joint data sets combined with BAO observations (Red line).
\label{fig6}}
\end{figure}

\section{Cosmological-model-independent constraints on compact source parameters}

In the previous section, we discussed the constraints on the model parameters of compact structure in radio sources using theoretical
expression for angular diameter distances and assuming best currently available $\Lambda$CDM cosmology.
In this section, we propose cosmological model independent approach.
Namely we derive the angular diameter distance for our radio sources from observed luminosity distance of SN Ia in the Union2 compilation
\citep{Amanullah10}. This provides us a natural way to
calibrate the properties of angular size of milliarcsecond radio
sources.

In order to place cosmological-model-independent constraints on $l$,
$\beta$, and $n$, one should first perform pairwise matching of
radio sources and SN Ia almost at the same redshift. Since the
sample size of the SN Ia is much larger than that of the radio
sources, we bin the observed $D_A$ (inferred from the Union2 data
points) according to the following criterion: $\Delta
z=|z_{radio}-z_{SN}|\leq 0.005$ \citep{Cao11a}. As a result we
obtain a sample of 42 observational angular diameter distances $D_A$
derived from the supernova data covering the redshift range
$0.031<z<1.34$. However, not all of them could be used in the
cosmological-model-independent method. As we already discussed in
Section 2 and Section 4, full $N=140$ sample is not statistically
complete and homogeneous. Moreover, sources corresponding to
different optical counterparts may have distinct ``angular size -
redshift" relation. Therefore, in cosmology independent analysis we
limited ourselves to radio sources with quasars as counterparts.
There are two reasons supporting this choice. First, our sample is
dominated by quasars. Second, as we have seen, cosmological
evolution of the linear size for quasars is very small, so we can
assume $n=0$. Therefore, we finally used SN Ia to derive
observational angular diameter distance for 26 quasars.

Fitting results of the compact source linear size parameters ($l$,
$\beta$) are shown in Fig.~\ref{fig5}, with the best fit
\begin{eqnarray}
&& l= 26.70\pm2.91 \ \mathrm{pc}, \nonumber\\
&& \beta= 0.158\pm 0.040. \nonumber
\end{eqnarray}
One can see that the results derived from the
cosmological-model-independent analysis agree very well with the
best-fitted parameters determined from theoretical cosmological
distances for the $N=112$ quasar sample.

Having performed cosmological-model-independent analysis, we can
consider cosmological implications of the corrected redshift -
angular size relation. Using the best-fitted $l$ and $\beta$
parameters (with their $1\sigma$ uncertainties) obtained from the
model-independent analysis to the full quasar data and performing
``angular size - redshift'' test assuming flat $\Lambda$CDM model,
we are able to get the observational constraint on the matter
density parameter. The result is $\Omega_m=0.292^{+0.065}_{-0.090}$
and its posterior probability density function is shown in
Fig.~\ref{fig6}. We see that it is in agrement with the value
obtained from Planck observations within 1$\sigma$ range around the
central value. Moreover, our analysis result is fully compatible
with that obtained from the previous study of peculiar velocities of
galaxies, $\Omega_m=0.30^{+0.17}_{-0.07}$, which is the only
alternative method sensitive exclusively to matter density
\citep{Feldman03}. Based on the 12 binned data-points with 12-13
compact sources per bin, it was found that Friedmann model with a
vanishing $\Lambda$ is not the best-fit cosmology
\citep{Vishwakarma01}. This result supported the necessity to
include dark energy in the cosmological model. Then \citet{Lima02}
used the same binned data to place constraints on the flat XCDM
cosmology (including dark energy with constant equation of state
$w$) with fixed physical length $l_m\simeq 20-30h^{-1}$ for the
radio sources. They demonstrated that the flat $\Lambda$CDM model
with $\Omega_m=0.20$ and $l_m=22.6h^{-1}$ is the best fit to these
milliarcsecond radio source data. The potential of using the same
sample to study other cosmological models including dark energy with
constant or time-variable equation of state was also discussed in
\citet{Chen03}. More recently, by applying an astrophysical model to
quantify the behavior of compact radio sources as standard rods and
considering possible selection effects, \citet{Jackson04} gave the
best-fit parameter $\Omega_m=0.24^{+0.09}_{-0.07}$ for the flat
$\Lambda$CDM model from the original data set \citep{Gurvits94}. We
find that the constraints resulting from our analysis are consistent
with the previous works. However, because we used the currently
favored cosmological model and performed a cosmological model
independent check, our results could be useful as hints for priors
on $l$, $\beta$ and $n$ parameters in future cosmological studies
using compact radio sources.

It has been known for some time that cosmological parameters can be
more stringently determined using additional and complementary data
\citep{Cao14}. Therefore, we combine the compact radio sources
``angular size - redshift'' test with the latest data on Baryon
Acoustic Oscillations (BAO). More specifically we add the BAO data
from the Sloan Digital Sky Survey (SDSS) data release 7 (DR7)
corresponding to $z=0.35$ \citep{Padmanabhan12}, SDSS-III Baryon
Oscillation Spectroscopic Survey (BOSS) at $z=0.57$
\citep{Anderson12}, the clustering of WiggleZ survey \citep{Blake12}
at $z=0.44, 0.60, 0.73$ and 6dFGS survey at $z=0.10$
\citep{Beutler11}. Likelihood distribution function for the
$\Omega_m$ parameter in $\Lambda$CDM model constrained by BAO and
the compact structure in quasars is also plotted in Fig.~\ref{fig6}.
Obviously, $\Omega_m$ is more tightly constrained with joint data
set, with the best-fit parameter $\Omega_m=0.308\pm0.019$. This also
agrees with that obtained from Planck observations very well.
Moreover, we find that cosmological constraint with the 112 quasars
is well consistent with the joint statistical analysis.

\section{Conclusion and discussion}

In this paper, we explored the properties of a sample of 140
milliarcsecond compact radio sources with measured angular sizes.
Metric linear size of compact sources is usually parameterized as
$l_m=l L^\beta (1+z)^n$. Using the best available cosmological model
parameters given by the Planck/WMAP9 observations, we investigated
the elements of $l_m$ --- its characteristic length $l$ as well its
dependence on the source luminosity $L$ and redshift $z$. In
the full sample, we found that measurements of $\theta$ provide
tighter estimate of the source linear size parameters. Small
cosmological evolution of the linear size $|n|\simeq 10^{-2}$ is
consistent with previous analyses, while a substantial evolution of
linear sizes with luminosity is still required ($\beta\simeq 0.17$).
However, the conclusion that $\beta+n\geq-0.15$ given by
\citet{Gurvits99} does not contradict to the findings of our work.

Furthermore, by dividing the full sample into seven different
sub-samples given the source redshifts and their optical
counterparts, we obtain the following results: (1) The range of the
parameters $\beta$ and $n$ for quasars are close to the the
estimates obtained for compact structures in BL Lac sources within
1$\sigma$. (2) Both luminosity dependence and weak redshift
dependence are present in radio galaxies. The best-fit values of
$\beta$ and $n$ for this sub-population are significantly different
from the corresponding quantities of quasars or BL Lac sources. (3)
Closeness or difference of parameter values for different types of
counterparts, might reveal the similar or different physical
processes governing the radio emission of compact structures. This
tendency could be also found from the constraints obtained with the
sub-samples located at different redshift bins: (1) Constraints on
the parameters with the low-redshit sub-sample ($z\leq 0.5$) are
essentially different from those obtained with other sub-samples.
(2) For the sub-sample with redshift range $0.5<z\leq 1.0$, the
``no-evolution'' model ($\beta=n=0$) is still included at $1\sigma$
confidence region in the $\beta - n$ plane, whereas a substantial
evolution of linear sizes with luminosity is still required for
other three sub-samples.

Finally, we studied the properties of angular size of milliarcsecond
radio quasars with a cosmological-model-independent method, and then
we derived the constraints (from the corrected quasar sample) on the
spatially flat $\Lambda$CDM cosmology. The obtained value of matter
density parameter, $\Omega_m=0.292^{+0.065}_{-0.090}$,
agrees very well with the previous results obtained on the
same ``$\theta$ - z'' sample and other recent astrophysical
measurements including Planck observations.

Therefore, our analysis indicates that, the radio source size seems
to be dependent on the source luminosity, i.e. the sources are not
``true'' standard rod. This is inconsistent with their model
previously discussed in the literature, in which this dependency has
been minimized by discarding low values of luminosities and extreme
values of spectral indices. However, in order to differentiate
observational selection effect from intrinsic luminosity-dependence,
we still need multi-frequency VLBI observations of more compact
radio sources with higher sensitivity and angular resolution.

As a final remark, we point out that the sample discussed in this
paper is based on VLBI images observed with various antennas
configurations and techniques for image reconstruction. Our analysis
potentially suffers from this systematic bias and taking it fully
into account will be included in our future work. Moreover, the
statistical results are obtained with VLBI images observed at
frequency of 5 GHz. Since the parameter $n$ may reflect possible
dependence of the linear size on the emitted frequency,
multi-frequency ``$\theta$ - z'' tests should also be included in
the future work.

\vspace{0.5cm}

The authors are grateful to the referee for very useful
comments which allowed to improve the paper. This work was
supported by the Ministry of Science and Technology National Basic
Science Program (Project 973) under Grants Nos. 2012CB821804 and
2014CB845806, the Strategic Priority Research Program ``The
Emergence of Cosmological Structure" of the Chinese Academy of
Sciences (No. XDB09000000), the National Natural Science Foundation
of China under Grants Nos. 11373014 and 11073005, the Fundamental
Research Funds for the Central Universities and Scientific Research
Foundation of Beijing Normal University, and China Postdoctoral
Science Foundation under grant No. 2014M550642. M.B. obtained
approval of foreign talent introducing project in China and gained
special fund support of foreign knowledge introducing project.

\end{document}